\documentclass[preprint,showpacs,preprintnumbers,amsmath,amssymb,nofootinbib]
{revtex4}

\usepackage{graphicx}% Include figure files
\usepackage{dcolumn}% Align table columns on decimal point
\usepackage{bm}% bold math
\usepackage{axodraw}
\usepackage{epsfig}

\newcommand{\be}{\begin{equation}}
\newcommand{\ee}{\end{equation}}
\newcommand{\bea}{\begin{eqnarray}}
\newcommand{\eea}{\end{eqnarray}}
\newcommand{\nn}{\nonumber}
\def\ve{\varepsilon}

\begin{document}

\begin{flushright}
%\preprint{hep-ph/xxyyzz}
%\today\\
%UH-0528
\end{flushright}

\preprint{}

\title{\large Implications of the non-universal $Z$ boson in FCNC mediated
rare decays}
\author{
R. Mohanta }
\affiliation{School of Physics, University of Hyderabad, Hyderabad - 500 046,
India}

\begin{abstract}

 We analyze the effect of the non-universal $Z$ boson in
the rare decays $B_s \to l^+ l^- $,
$B_s \to l^+ l^- \gamma $ and $Z \to b \bar s$ decays. These
decays involve  the  FCNC mediated $ b \to s $ transitions, and are found
to be very small in the standard model. The smallness of these decays
in the standard model makes them sensitive probe for new physics. We find
an enhancement of at least an order in these branching ratios because of
the non-universal $Zbs$
coupling.

\end{abstract}

\maketitle

The rare decays induced by the flavor changing neutral current (FCNC)
transitions are very important to probe the flavor sector of the
standard model (SM). In the SM they arise from one-loop diagrams
and hence are generally suppressed in comparison to the tree diagrams.
This makes them as a sensitive probe for new physics.
The rare $B$ decays which are mediated by the FCNC transitions are of the
kind $b \to s $ or $b \to  d$. Prominent examples
of rare $B$ decays are $B \to K^* \gamma$, $B \to \rho \gamma$,
$B \to K l^+ l^-$, $B_{s,d} \to l^+ l^-$. During the last few years,
considerable theoretical attention has therefore been focused on
these decays  in view of the planned
experiments at B-factories, which are likely to measure branching
fractions as low as $10^{-8}$ \cite{ref1,ref0}. The results of the branching
ratios of FCNC mediated $B$-decays are very sensitive to new physics beyond
the SM. Thus, a detailed investigation of the rare decays is a promising way
to discover or severely constrain the new physics.

In this paper, we focus on a specific class
of  rare decay modes $B_s \to l^+ l^- $ and
$B_s \to l^+ l^- \gamma $, which are mediated
by the $Z$ boson exchange and the rare  $Z$ decay mode $Z \to b \bar s $.
We consider the effect of the non-universal $Z$ boson which induces
FCNC interaction at the tree level. It is well known that
FCNC coupling of the $Z$ boson can be generated at the tree
level in various exotic scenarios. Two popular examples discussed in the
literature are the models with an extra $U(1)$ symmetry
\cite{ref2} and those with the addition of non-sequential generation
of quarks \cite{ref3}. In the case of extra $U(1)$ symmetry the FCNC
couplings of the $Z$ boson are induced by $Z -
Z^\prime$ mixing, provided the SM quarks have family non-universal
charges under the new $U(1)$ group. In the second case, adding a different
number of up- and down-type quarks, the pseudo CKM matrix
needed to diagonalize the charged currents is no longer unitary
and this leads to tree level FCNC couplings. It should be 
noted that, recently, there 
has been renewed interests shown in the literature concering the
non-universal $Z$ induced new physics \cite{barger}. In light of this
it necessitates also a detailed investigation of rare $B$ decays, which are
very promising to discover and/or to constrain new physics. 

Here we will follow the second
approach \cite{ref3} to analyze some  FCNC induced rare decays.
It is a simple model beyond
the standard model with an enlarged matter sector due to an additional
vector like down quark $D_4$. The presence of an additional
down quark implies a $4 \times 4$ matrix $V_{i \alpha}$ $(i=u,c,t,4,
 ~ \alpha=d,s,b, b^\prime)$, diagonalizing the down quark
mass matrix. For our purpose the relevant information for
the low energy physics is encoded in the extended mixing matrix.
The charged currents are unchanged except that the $V_{CKM}$ is now
the $3 \times 4$ upper sub-matrix of $V$. However, the distinctive
feature of this model is that the FCNC interaction enters neutral current
Lagrangian of the left handed down quarks as
\be
{\cal L}_Z= \frac{g}{2 \cos \theta_W}\Big[\bar u_{Li}
\gamma^\mu u_{Li}-\bar d_{L \alpha}U_{\alpha \beta} \gamma^\mu
d_{L \beta}-2 \sin^2 \theta_W J^\mu_{em}\Big] Z_\mu\;,
\ee
with
\be
U_{\alpha \beta}= \sum_{i=u,c,t} V_{\alpha i}^\dagger V_{i \beta}=
\delta_{\alpha \beta}-V_{4 \alpha}^* V_{4 \beta}\;,
\ee
where $U$ is the neutral current mixing matrix for the
down sector, which is given above. As $V$ is not unitary,
$U \neq {\bf 1}$. In particular the non-diagonal
elements do not vanish.
\be
U_{\alpha \beta}=-V_{4 \alpha}^* V_{4 \beta} \neq 0 ~~~~
{\rm for}~~~\alpha \neq \beta\;.
\ee
Since the various $U_{\alpha \beta}$ are non vanishing, they would
signal new physics and the presence of FCNC at the tree level and this can
substantially modify the predictions of SM for the FCNC processes.

Now let us consider the FCNC process $B_s \to
l^+ l^-$. These decays, in particular the process $B_s \to
\mu^+ \mu^-$ has attracted a lot of
attention recently since it is very sensitive to the structure of SM and
potential source of  new physics beyond the SM. Furthermore, this process
is very clean and the only nonperturbative quantity involved is
the decay constant of
$B_s$ meson which can be reliably calculated by the well known
non-perturbative methods such as QCD sum rules, lattice gauge theory etc.
Therefore, it provides a good hunting ground to probe for new physics.
The branching ratio for $B_s \to l^+ l^-$ has been calculated in the
SM \cite{ref4} and also in beyond the SM in a number of papers \cite{ref5}.
Let us start by recalling the result for $B_s \to l^+ l^-$
in QCD-improved standard model. The effective Hamiltonian
describing this process is
\bea
\label{one}
{\cal{H}}_{eff}
&=& \frac{G_F \alpha }{\sqrt{2} \pi} V_{tb} V_{ts}^*
\Bigg[
C^{eff}_9 ~({\bar s}~ \gamma_\mu~ P_L ~b)
({\bar l}~ \gamma^\mu ~l)
+ C_{10}~({\bar s}~ \gamma_\mu~ P_L ~b)({\bar l} ~\gamma^\mu ~\gamma_5 ~l)
\nonumber \\
&&
~~~~~~~- \frac{2 C_7~ m_b}{q^2} ({\bar s}i \sigma_{\mu \nu}
q^\nu P_R ~b)
( {\bar l}~\gamma^\mu ~ l)
\Bigg]\;,\label{ham}
\eea
where
$P_{L,R} = \frac{1}{2}~(1 \mp \gamma_5)$ and $q$ is the momentum
transfer. 
$C_i$'s are the Wilson coefficients evaluated at the $b$ quark mass 
scale in NLL order with values \cite{beneke} 
\be
C_7^{eff}=-0.308\;,~~C_9=4.154\;,~~C_{10}=-4.261\;.\label{wil}
\ee
The coefficient $C_9^{eff}$ has a perturbative part and a 
resonance part which comes
from the long distance effects due to the conversion of the real
$c \bar c$ into the lepton pair $l^+ l^-$. Hence, $C_9^{eff}$ can be 
written as 
\be
C_9^{eff}=C_9+Y(s)+C_9^{res}\;,
\ee
where the function $Y(s)$ denotes the perturbative part coming 
from one loop matrix elements  of the four quark operators and
is given in Ref. \cite{ab}.
The long distance resonance effect is given as \cite{res}
\bea
C_9^{res}= \frac{3 \pi}{\alpha^2}(3 C_1+C_2+3C_3+C_4+3C_5+C_6)\sum_{J/\psi,
\psi^\prime} \kappa\frac{m_{V_i} \Gamma(V_i \to l^+ l^-)}{m_{V_i}^2 -s
-i m_{V_i}\Gamma_{V_i}}\;,
\eea
where the phenomenological parameter $\kappa$ is taken to be 2.3, so as to
reproduce the correct branching ratio  $ {\cal B}(B \to J/\psi K^*
\to K^* l^+ l^-)={\cal B}(B \to J/\psi K^*){\cal B}(J/\psi \to l^+ l^-)$.
In this analysis, we will consider only 
the contributions arising from two dominant resonances i.e., $J/\psi$
and $\psi^\prime$. The values of the
coefficients $C_i$'s in NLL order are given in \cite{beneke} as
$C_1=-0.151\;,~C_2=1.059\;,~C_3=0.012\;,~C_4=-0.034\;,~
C_5=0.010$ and $C_6=-0.040$. 

To evaluate the transition amplitude one can generally adopt the vacuum
insertion method, where the form factors of the various currents
are defined as follows
\bea
&&\langle 0~|~ {\bar s} ~\gamma^\mu~ \gamma_5 ~b~| B^0_s \rangle
= i f_{B_s} p^\mu_B\;, \nn\\
&&\langle 0~| {\bar s} ~\gamma_5~ b| B^0_s \rangle
=  i f_{B_s} m_{B_s}\;,\nn\\
&&\langle 0 |~ {\bar s} ~\sigma^{\mu \nu} ~P_R ~b~ |B^0_s\rangle = 0\;.
\label{four}
\eea
Since $ p^\mu_B = p^\mu_+ + p^\mu_-$, the contribution from
$C_9$ term in Eq. (\ref {one})
will vanish upon contraction with the lepton bilinear, $C_7$
will also give zero
by (\ref{four}) and the remaining $C_{10}$ term will get a factor of
$2m_l$.
\par Thus the transition amplitude for the process is given as
\bea
\label{four1}
{\cal M}(B_s \to l^+ l^-)
&=& i\frac{G_F~ \alpha }{\sqrt{2} \pi} ~V_{tb} V_{ts}^*~ 
f_{B_s}~ C_{10}~ m_l
~ (\bar{l} \gamma_5 l)\;,
\eea
and the corresponding branching ratio is given as
\be
{\cal B}(B_s \to l^+ ~l^-)
= \frac{ G_F^2~ \tau_{B_s}}{16 \pi^3}~\alpha^2~ f_{B_s}^2~ m_{B_s}~ m_l^2
~|V_{tb}V^*_{ts}|^2 ~C_{10}^2 ~\sqrt{1-\frac{4 m_l^2}{m_{B_s}^2}}\;.
\label{five}
\ee
Helicity suppression is reflected by the presence of $m_l^2$ in
(\ref{five}) which gives almost vanishingly small value
for $e^+ e^-$ and  a very small
branching ratio of $(3.4 \pm 0.5)\times
10^{-9}$ for $\mu^+ \mu^-$ \cite{buras}. The  published Tevatron/CDF
physics results with luminosity $171~{\rm pb}^{-1}$
provides the bound on
$ B_s \to \mu^+ \mu^-$ \cite{cdf}
\be
{\cal B}(B_s \to \mu^+ \mu^-) < 5.8 \times 10^{-7}\,~~~~~{( 90\%~
{\rm C.L.})}\;.
\ee
Recently, this branching ratio has been further constrained by the  D0 
collaboration \cite{d0} with an upper bound
\be
{\cal B}(B_s \to \mu^+ \mu^-) < 5.0 \times 10^{-7}\,~~~~~{( 95\%~
{\rm C.L.})}\;.
\ee
It should be noted that the $\tau $ channel is free from this helicity 
suppression however,
its experimental detection is quite hard due to the low detection
 efficiency and
that is why we do not have any  experimental upper limit 
for this process as yet.

\hspace*{1.0 truein}
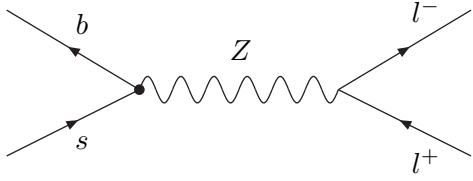
\begin{figure}[t]
\begin{center}
\begin{picture}(300,50)(0,0)
\SetColor{Black}
\Photon(100,25)(175,25){5}{6}
\SetColor{Black}
\ArrowLine(50,0)(100,25){}{}
\SetColor{Black}
\ArrowLine(100,25)(50,55){}{}\Vertex(100,25){2}
\SetColor{Black}
\ArrowLine(225,0)(175,25){}{}
\SetColor{Black}
\ArrowLine(175,25)(225,55){}{}

\vspace*{0.8 true in}
%\Text(130,-5)[]{(a)}
\Text(135,40)[]{$Z$}
\Text(75,50)[]{$b$}
\Text(75,05)[]{$s$}
\Text(205,55)[]{$l^-$}
\Text(205,-02)[]{$l^+$}

\end{picture}
\end{center}
\caption{Feynman diagram for $B_s \to l^+ l^-$ in the model with tree
level FCNC transitions, where the blob
represents the tree level flavor changing vertex.}
\end{figure}

Now let us analyze the decay modes $B_s \to l^+ l^-$ in the model with the
$Z$ mediated FCNC
occurring at the tree level \cite{ref3}. The corresponding diagram
is shown in Figure-1, where the blob represents the tree level $Zbs$
coupling. The effective Hamiltonian for $B_s \to l^+ l^-$ is given as
\be
{\cal H}_{eff}= \frac{G_F}{\sqrt 2}~ U_{sb}~ [\bar s \gamma^\mu
(1-\gamma_5)b]
\left [\bar l( C_V^l \gamma_\mu  -C_A^l \gamma_\mu \gamma_5 ) l
\right ]\;,\label{ham1}
\ee
where $C_V^l$ and $C_A^l$  are the vector and axial vector $Z l^+ l^-$
couplings, which are given as
\be
C_V^l= -\frac{1}{2}+2\sin^2 \theta_W\;,~~~~~~~
C_A^l= -\frac{1}{2}\;. \label{ca}
\ee
Since, the structure of the effective Hamiltonian (\ref{ham1}) in
this model is the same form  as that of the SM, like $\sim(V-A)(V-A)$ form,
therefore its 
 effect on the various decay observables can be encoded by replacing the 
SM Wilson coeiffients $(C_9^{eff})^{SM}$ and $(C_{10})^{SM}$ by 
\bea
C_9^{eff} =( C_9^{eff})^{SM}+\frac{2 \pi}{\alpha}\frac{ U_{sb} C_V^l}{
V_{tb}V_{ts}^*}\;,~~~~
C_{10}^{eff} = (C_{10})^{SM}-\frac{2 \pi}{\alpha}\frac{ U_{sb}C_A^l}{
V_{tb}V_{ts}^*}\;.\label{nph}
\eea
It should be noted that $U_{sb}$ is in general complex and hence  it
induces the weak phase difference ($\theta$) between the SM and new 
physics contributions. Since the value of the Wilson coefficients $C_9$ 
and $C_{10}$ are opposite to each other as seen from Eq. (\ref{wil}),
and  the  new physics contributions to $C_9$ and $C_{10}$ are opposite 
to each other, one will get constructive or destructive interference 
of SM and NP amplitudes for $\theta=\pi$ or zero (where $\theta$ denotes the
relative weak phase between SM and NP contribution in the above equation). 
 
Thus, one can obtain the branching ratio including the NP contributions by 
substituting $C_{10}^{eff}$ from (\ref{nph}) in (\ref{five}). 
Now using the value of $|U_{bs}| \simeq 10^{-3}$ \cite{ref6}, which has been
extracted from the recent data on ${\cal B}(B \to X_S l^+
l^-)$, $\sin^2 \theta_W=0.23$, the particle masses 
from \cite{pdg}, $\alpha=1/127$, the decay constant
$f_{B_s}=0.24$ GeV and $V_{tb} V_{ts}^*=0.04$, we obtain 
the branching ratios as
\bea
{\cal B}(B_s \to \mu^+ \mu^-) &=& 4.2 \times 10^{-8}\;,~~~({\rm for}~
\theta= \pi)\nn\\
&=& 6.8 \times 10^{-9}\;,~~~({\rm for}~
\theta= 0)\;,\nn\\
{\cal B}(B_s \to \tau^+ \tau^-) &=& 8.9 \times 10^{-6}\;,~~~({\rm for}~
\theta= \pi)\nn\\
&=& 1.4 \times 10^{-6}\;,~~~({\rm for}~
\theta= 0)\;.
\label{eq1}
\eea
Thus, as seen from Eq. (\ref{eq1}) the branching ratio for
$B_s^0 \to \mu^+ \mu^-$
has been  enhanced  by one order from its corresponding standard model value
for $\theta=\pi$,
and is below the present experimental upper limit.  This decay mode may
be observable at the Tevatron Run II \cite{tev} to the level of
$2 \times 10^{-8}$. However, the predicted
branching ratio for  $B_s^0 \to \tau^+ \tau^-$, which is ${\cal O}(10^{-6})$,
could be observable in the currently running $B$ factories, if we
have a good efficiency for the detection of $\tau$ lepton.

Now let us consider the radiative dileptonic decay modes
$B_s \to l^+ l^- \gamma $, which are also very sensitive to the
existence of new physics beyond the SM. Due to the presence
of the photon in the final state, these decay modes are free from
helicity supression, but they are further suppressed by a
factor of $\alpha $. However, in spite of this
$\alpha $ suppression,  the radiative leptonic decays $B_s \to 
l^+ l^- \gamma $,
$l=(\mu, \tau)$  have comparable decay rates to that of
purely leptonic ones. The SM predictions for these
branching ratios are
${\cal B}(B_s \to \mu^+ \mu^- \gamma,~\tau^+ \tau^- \gamma)=1.9 \times
10^{-9},~9.54 \times 10^{-9}$ respectively  \cite{eilam, smg}. 
These decays are also studied in some
beyond the stanadard model scenarios \cite{aliev}.

The matrix element for the decay $B_s \to l^+ l^- \gamma$ can
be obtained from that of the  $B_s \to l^+ l^-$ one  by attaching the photon
line to any of the charged external fermion lines. In order to
calculate the amplitude, when the photon is
radiated from the initial fermions (structure dependent (SD) part), we
need to evaluate the  matrix elements of the quark currents present
in (\ref{ham}) between the emitted  photon and the initial $B_s$
meson. These matrix elements can be obtained by considering the
transition of a $B_s$ meson to a virtual photon with momentum $k$.
In this case the form factors depend on two variables, i.e., $k^2$ (the
photon virtuality) and the square of momentum transfer $q^2=(p_B-k)^2$.
By imposing gauge invariance, one can obtain several relations
among the form factors at $k^2=0$. These relations can be used 
to reduce the number of independent form factors for the transition of 
the $B_s$ meson to a real photon. Thus, the matrix elements for $B_s
\to \gamma$ transition, induced by vector, axial-vector, tensor and 
pseudotensor currents can be parametrized as \cite{kruger}
\bea
\langle \gamma(k, \ve)|\bar s \gamma_\mu \gamma_5 b|B_s(p_B) \rangle
&=& ie \left [ \ve_\mu^* (p_B\cdot k) -(\ve^* \cdot p_B) k_\mu \right ]
\frac{F_A}{m_{B_s}}\;,\nn\\
\langle \gamma(k, \ve)|\bar s \gamma_\mu  b|B_s(p_B) \rangle
&=& e\epsilon_{\mu \nu \alpha \beta} \ve^{*\nu} p_B^\alpha~ k^\beta
\frac{F_V}{m_{B_s}}\;,\nn\\
\langle \gamma(k, \ve)|\bar s \sigma_{\mu \nu} q^\nu 
\gamma_5b|B_s(p_B) \rangle
&=& e \left [ \ve_\mu^* (p_B\cdot k) -(\ve^* \cdot p_B) k_\mu \right ]
F_{TA}\;,\nn\\
\langle \gamma(k, \ve)|\bar s \sigma_{\mu \nu} q^\nu b|B_s(p_B) \rangle
&=& e\epsilon_{\mu \nu \alpha \beta} \ve^{*\nu} p_B^\alpha~ k^\beta
F_{TV}\;,
\eea
where $\varepsilon$ and $k$ are the polarization vector and the
four-momentum of photon, $p_B$ is the momentum of initial $B_s$
meson and $F_i$'s are the various form factors.  

Thus, the matrix element describing the SD part takes the form
\bea
{\cal M}_{SD} &=& \frac{\alpha^{3/2}G_F}{\sqrt{2 \pi}}~ 
V_{tb}V_{ts}^*
 \biggr\{ \epsilon_{\mu \nu \alpha \beta}
\varepsilon^{* \nu} p_B^\alpha~ k^\beta\Big(A_1~ \bar l \gamma^\mu l
+A_2~ \bar l \gamma^\mu \gamma_5 l \Big)\nn\\
&+&
i\Big( \varepsilon_\mu^*(k \cdot p_B)-(\varepsilon^* \cdot p_B) k_\mu
\Big)\Big(B_1~ \bar l \gamma^\mu l
+B_2~ \bar l \gamma^\mu \gamma_5 l \Big)\biggr\}\;,
\label{sd}
\eea
where
\bea
A_1&=& 2 C_7 \frac{m_b}{q^2}F_{TV}+C_9 \frac{F_V}{m_{B_s}}\;,~~~
~~~~~~~~~~A_2=C_{10}\frac{F_V}
{m_{B_s}}\;,\nn\\
B_1&=& -2C_7\frac{m_b}{q^2} F_{TA}-C_9 \frac{F_A}{m_{B_s}}\;,~~~
~~~~~~~~~B_2=-C_{10} \frac{F_A}{m_B}\;.\label{ff}
\eea
The  $q^2$ dependence of the
form factors are given as \cite{kruger}
\be
F(E_\gamma)= \beta \frac{f_{B_s} m_{B_{s}}}{\Delta+ E_\gamma}\;,
\label{ib}
\ee
where $E_\gamma$ is the photon energy, which is related to the 
momentum transfer $q^2$ as
\be
E_\gamma= \frac{m_{B_s}}{2}\left (1- \frac{q^2}{m_{B_s}^2} \right )\;.
\ee
The values of the parameters are given in Table-1.
%%%%%%%%%%%%%%%%%%%%%%%%%%%%%%%%%%%%%%%%%%%%%%%%%%%%%%%%%%%%%%%%
\begin{table}
\begin{center}
\caption{The parameters for $B \to \gamma$ form factors.}
\vspace*{0.3 true cm}
\begin{tabular}{|c|cc|c c|cc|cc|}
\hline
\hline
Parameter & & $ F_V $ && $ F_{TV}$ && $F_A$ && $F_{TA}$ \\
\hline
$\beta ({\rm GeV}^{-1})$ && 0.28  && 0.30 && 0.26 && 0.33 \\
$\Delta ({\rm GeV})$ && 0.04  && 0.04 && 0.30 && 0.30 \\
\hline
\hline
\end{tabular}
\end{center}
\end{table}
%%%%%%%%%%%%%%%%%%%%%%%%%%%%%%%%%%%%%%%%%%%%%%%%%%%%%%%%%%%%%%%%%%%%%%

When the photon is radiated from the outgoing lepton pairs, the
internal bremsstrahlung  (IB) part, the matrix
element is given as \cite{gaur}
\be
{\cal M}_{IB} = \frac{\alpha^{3/2}G_F}{\sqrt{ 2 \pi}}~
V_{tb}V_{ts}^*~  f_{B_s}~ m_l~ C_{10}
\biggr[ \bar l \left ( \frac{\not\!{\varepsilon}^* {\not\!{p}}_{B}}{
p_+ \cdot k}-\frac{{\not\!{p}}_{B}\not\!{\varepsilon}^* }{
p_- \cdot k} \right )\gamma_5 ~l \biggr]\;.
\ee
Thus, the total matrix element for the $B_s \to l^+ l^- \gamma $ process
is given as
\be
{\cal M}={\cal M}_{SD}+{\cal M}_{IB}\;.
\ee
The differential decay width of the $B \to l^+
l^- \gamma $ process, in the rest frame of $B_s$ meson is given as \cite{gaur}
\be
\frac{d \Gamma}{d s}= \frac{G_F^2 \alpha^3}{2^{10} \pi^4}~ |V_{tb} 
V_{ts}^*|^2~
m_{B_s}^3~ \Delta\;,\label{lp}
\ee
where
\bea
\Delta &=& \frac{4}{3} m_{B_s}^2 (1- \hat s)^2 v_l \Big((\hat s+2r_l)(|A_1|^2
+|B_1|^2)+(\hat s-4 r_l)(|A_2|^2+|B_2|^2 \Big )\nn\\
&-& 64~ \frac{f_{B_s}^2}{m_{B_s}^2} \frac{r_l}{1- \hat s}~ C_{10}^2~\Big(
(4r_l-\hat s^2 -1) \ln
\frac{1+v_l}{1-v_l}+2 \hat s~ v_l\Big)\nn\\
&-& 32~r_l(1-\hat s)^2~ f_{B_s} {\rm Re}\Big( C_{10} A_1^* \Big),
\eea
with  $s=q^2$, $\hat s= s/m_{B_s}^2$, $r_l=m_l^2/m_{B_s}^2$,
$v_l=\sqrt{1- 4 m_l^2/q^2}$. The physical region of $s$ is 
$4 m_l^2 \leq s \leq m_{B_s}^2 $.

The forward backward asymmetry is given as
\bea
A_{FB} &=& \frac{1}{\Delta} \biggr[2 m_{B_s}^2 \hat s(1-\hat s)^3 v_l^2~
{\rm Re}\Big(
A_1^* B_2+B_1^* A_2\Big)\nn\\
&+&32~ f_{B_s}~r_l (1-\hat s)^2 \ln\left (\frac{4r_l}{\hat s} \right )
{\rm Re}\Big(C_{10}B_2^*\Big)\biggr]\;.\label{fb}
\eea 
Now using the form factors from (\ref{ff}),
we plot the dilepton mass spectrum (\ref{lp}),
and the forward backward asymmetries (\ref{fb}) for $B_s \to l^+ l^- \gamma$ 
decays which are shown in Figures-2 and 3. In these plots we have used the 
weak phase difference  between the SM and NP amplitudes $\theta$
to be $\pi$ to get the maximum possible contributions. From figures-2 and 3,
we see that the branching ratio for $B_s \to l^+ l^- \gamma $ enhanced
significantly from their corresponding SM values. However, the forward 
backward asymmetries are reduced slightly from the corresponding SM values
and for the $B_s \to \mu^+ \mu^- \gamma$ process, there is a backward 
shifting of the zero position.  
 %%%%%%%%%%%%%%%%%%%%%%%%%%%%%%%%%%%%%%%%%%%%%%%%%%%%%%%%%%%%%%%%%%%%%%
\begin{figure}[htb]
   \centerline{\epsfysize 3.0 truein \epsfbox{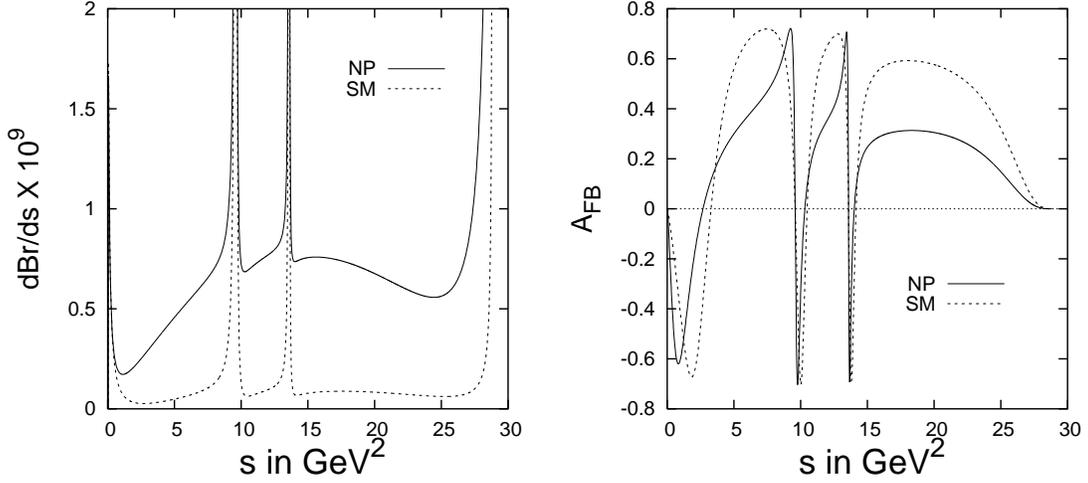}}
 \caption{
  The differential branching ratio and the forward backward 
asymmetry ($A_{FB}$) for the process  
$B_s \to   \mu^+ \mu^- \gamma $, in the standard model and in
the  NP model with  the non-universal 
$Z$-boson effect.}
  \end{figure}
%%%%%%%%%%%%%%%%%%%%%%%%%%%%%%%%%%%%%%%%%%%%%%%%%%%%%%%%%%%%%%%%%%%%
\begin{figure}[htb]
   \centerline{\epsfysize 3.0 truein \epsfbox{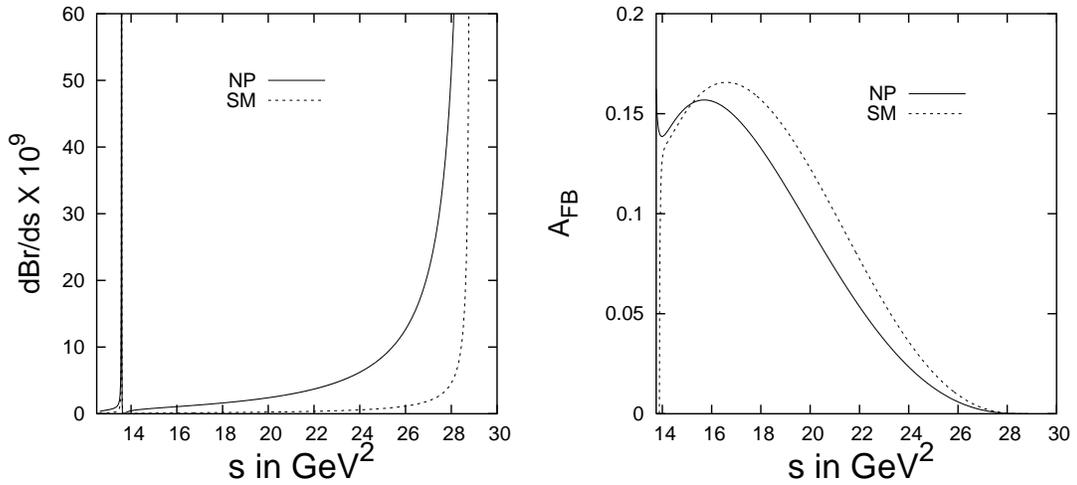}}
 \caption{Same as Figure-2, for the  $B_s \to   \tau^+ \tau^- \gamma $
process.  }
  \end{figure}
%%%%%%%%%%%%%%%%%%%%%%%%%%%%%%%%%%%%%%%%%%%%%%%%%%%%%%%%%%%%%%%%%%%%%

To obtain the branching ratios it is necessary
to eliminate the backgrounds, coming from the resonances  $J/\psi 
(\psi^\prime)$
with $J/\psi(\psi^\prime) \to l^+ l^- $. We use the
following  veto
windows to eliminate these backgrounds
\begin{eqnarray*}
B_s \to \mu^+ \mu^- \gamma :&&m_{J/\psi}-0.02<
m_{\mu^+ \mu^-}<m_{J/\psi}+0.02;\nn\\
:&&
 m_{\psi^\prime}-0.02<m_{\mu^+ \mu^-}<m_{\psi^\prime}+0.02 \nn\\
B_{s} \to  \tau^+ \tau^- \gamma:&&
m_{\psi^\prime}-0.02<m_{\tau^+ \tau^-}<m_{\psi^\prime}+0.02 \;.
\end{eqnarray*}
Furthermore, it should be noted that the 
$|{\cal M}_{IB}|^2$ has infrared singularity
due to the emission of soft photon. Therefore, to obtain the branching ratio,
we impose a cut on the photon energy, which will correspond to the
experimental cut imposed on the minimum energy for the detectable photon.
Requiring the photon energy to be larger than 25 MeV, i.e.,
$E_\gamma \geq \delta~ m_{B_s}/2$, which corresponds to
$s \leq m_{B_s}^2(1- \delta)$ and therefore, we set 
the cut $\delta \geq 0.01 $.

Thus, with the above defined veto windows and the infrared cutoff parameter,
 we obtain the brancing ratios as
\bea
{\cal B}(B_s^0 \to \mu^+ \mu^- \gamma ) &=& 1.94 \times 10^{-8}\;,
\nn\\
{\cal B}(B_s^0 \to \tau^+ \tau^- \gamma) &=& 1.37 \times 10^{-7}\;,
\eea
which are enhanced by an order from their SM values.
It should be mentioned that the $B_s^0 \to \tau^+ \tau^- \gamma$
could be observable in the Run II of Tevatron. The contribution to the 
branching ratio due to bremsstrahlung photon
is small for $B_s \to \mu^+ \mu^- \gamma$ and is found to be
${\cal B}(B_s^0 \to \mu^+ \mu^- \gamma )|_{IB} \sim 0.5 \times 10^{-8}$
whereas it has dominant contribution to the $B_s \to \tau^+ \tau^- 
\gamma$ process,
i.e.,  ${\cal B}(B_s^0 \to \tau^+ \tau^- \gamma)|_{IB} \sim 1.3 \times 
10^{-7}$.

Now let us consider the flavor changing rare $Z$ decays
$Z \to b \bar s $. Rare $Z$ decays have been studied extensively
in order to yield the signature of new physics.
In the standard model this mode originates from
one loop diagram with branching ratio $\sim 3 \times 10^{-8}$ \cite{sm}.
While the sensitivity of the measurement
for the branching ratios for rare $Z$ decays reached at LEP2
is about $10^{-5}$ \cite{pdg}, future linear colliders (NLC, TESLA)
will bring this sensitivity up to $10^{-8}$ level \cite{agular}.
Various
beyond the standard model scenarios has been employed \cite{bsm} where
the branching ratio can be found to reach the sensitivity of the
order of  $10^{-6}$. We would now be interested to analyze this decay mode
in the model with an additional vector like down quark,
where it can originate in the leading order.
For completenes, we would also include the one loop corrections
to the branching ratio, although
their effect is negligibly small compared to the tree level contribution.

%%%%%%%%%%%%%%%%%%%%%%%%%%%%%%%%%%%%%%%%%%%%%%%%%%%%%%%%%%%%%%%%%%%%%%%
\hspace*{1.0 truein}
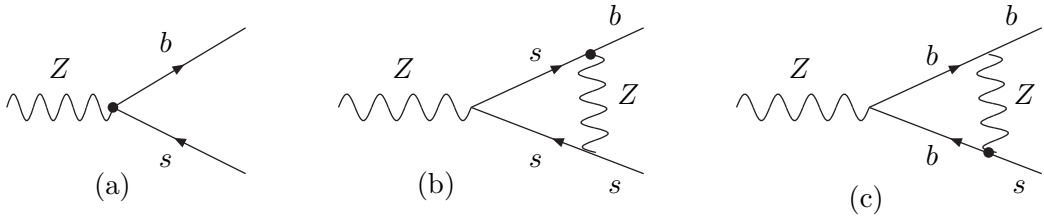
\begin{figure}[b]
%\begin{boldmath}
\begin{center}
\begin{picture}(500,60)(-45,0)
\SetColor{Black}
\Photon(-25,25)(15,25){5}{4}
\Photon(100,25)(150,25){5}{4}
\Photon(250,25)(300,25){5}{4}
\SetColor{Black}
\ArrowLine(65,0)(15,25){}{}
\SetColor{Black}
\ArrowLine(15,25)(65,55){}{}\Vertex(15,25){2}
\SetColor{Black}
\ArrowLine(215,0)(150,25){}{}
\SetColor{Black}
\ArrowLine(150,25)(215,55){}{}\Vertex(195,45){2}
\Photon(197,8)(195,45){5}{4}

\ArrowLine(365,0)(300,25){}{}
\ArrowLine(300,25)(365,55){}{}\Vertex(345,8){2}
\Photon(345,45)(348,8){5}{4}
\vspace*{0.8 true in}
\Text(130, -10)[lb]{(b)}
\Text(15,-5)[]{(a)}
\Text(300,-10)[]{(c)}
\Text(-05,40)[]{$Z$}
\Text(125,40)[]{$Z$}
\Text(275,40)[]{$Z$}
\Text(35,50)[]{$b$}
\Text(35,05)[]{$s$}
\Text(205,60)[]{$b$}
\Text(205,-05)[]{$s$}
\Text(175,45)[]{$s$}
\Text(175,05)[]{$s$}
\Text(210,30)[]{$Z$}
\Text(360,30)[]{$Z$}
\Text(325,45)[]{$b$}
\Text(325,07)[]{$b$}
\Text(355,60)[]{$b$}
\Text(358,-05)[]{$s$}
\end{picture}
\end{center}
%\end{boldmath}
\caption{Tree-level and one loop Feynman diagrams for $Z \to b \bar s $
process.}
\end{figure}

%%%%%%%%%%%%%%%%%%%%%%%%%%%%%%%%%%%%%%%%%%%%%%%%%%%%%%%%%%%%%%%%%%%%%%%%%%%
The tree level amplitude is given as
\be
{\cal M}(Z \to b \bar s)= \epsilon^\mu~ U_{sb} ~\frac{g}{2 \cos \theta_W}
\bar u_b~ \gamma_\mu~ P_L~v_s\;,
\ee
where $\epsilon^\mu$ denotes the polarization vector of $Z$ boson.
The decay width is is given as
\be
\Gamma(Z \to b \bar s)= \frac{m_Z}{32 \pi}~ \frac{g^2}{\cos^2 \theta_W}~
|U_{sb}|^2\;,
\ee
and the branching ratio to be 
\be
{\cal B}(Z \to b \bar s+\bar b s)=\frac{1}{\Gamma_Z}~\frac{m_Z}{16 \pi}
~\frac{g^2}{\cos^2 \theta_W}|U_{sb}|^2\;,
\ee
where $\Gamma_Z$ is the total $Z$-boson decay width.

Now we consider
the one loop corrections arising from Figures 4-(b) and (c).
The one-loop amplitude arising from Figure-4 (b) is given as
\bea
{\cal M}(Z \to b \bar s)|_{1-loop} &=& \frac{1}{16 \pi^2}
\epsilon^\mu U_{sb}
~\frac{g}{2 \cos \theta_W}~\bar u_b~ a_L^2 \gamma_\mu~ P_L
\biggr [ -2 \tilde C_0(m_Z^2, m_s^2,m_s^2,m_b^2, m_Z^2,m_s^2)\nn\\
& +&B_1 (m_s^2, m_s^2, m_Z^2)+B_0(m_s^2, m_s^2, m_Z^2)
 -\tilde C_{11}(m_Z^2, m_s^2,m_s^2,m_b^2, m_Z^2,m_s^2)\nn\\
&+&2C_{24}(m_Z^2, m_s^2,m_s^2,m_b^2, m_Z^2,m_s^2)
+m_Z^2\biggr(2C_{11}(m_Z^2, m_s^2,m_s^2,m_b^2, m_Z^2,m_s^2)\nn\\
&+&3C_0(m_Z^2, m_s^2,m_s^2,m_b^2, m_Z^2,m_s^2)
-C_{22}(m_Z^2, m_s^2,m_s^2,m_b^2, m_Z^2,m_s^2)\nn\\
&+&C_{23} (m_Z^2, m_s^2,m_s^2,m_b^2, m_Z^2,m_s^2)\biggr)\biggr ] v_s\;,
\eea
where
\be
a_L=\frac{ g}{\cos \theta_W}\left (\frac{1}{3}\sin^2 \theta_W- \frac{1}{2}
\right )\;.
\ee
The contribution from Figure-4 (c) can be obtained from 4-(b) by replacing
$m_s$ by $m_b$ and vice-versa.
Including the one-loop correction the branching ratio for $Z \to (b \bar
s + \bar b s ) $ is given as
\be
{\cal B}(Z \to b \bar s+\bar b s)=\frac{1}{\Gamma_Z}~\frac{m_Z}{16 \pi}
~\frac{g^2}{\cos^2 \theta_W}|U_{sb}|^2 |1+R_1+R_1(m_s \leftrightarrow
m_b)|^2\;,
\ee
where
\bea
R_1 &=& \frac{a_L^2}{16 \pi^2} \biggr [ -2 \tilde C_0
(m_Z^2, m_s^2,m_s^2,m_b^2, m_Z^2,m_s^2)
 +B_1 (m_s^2, m_s^2, m_Z^2)\nn\\
&+&B_0(m_s^2, m_s^2, m_Z^2)
 -\tilde C_{11}(m_Z^2, m_s^2,m_s^2,m_b^2, m_Z^2,m_s^2)
+2C_{24}(m_Z^2, m_s^2,m_s^2,m_b^2, m_Z^2,m_s^2)\nn\\
&+&m_Z^2\biggr(2C_{11}(m_Z^2, m_s^2,m_s^2,m_b^2, m_Z^2,m_s^2)
+3C_0(m_Z^2, m_s^2,m_s^2,m_b^2, m_Z^2,m_s^2)\nn\\
&-&C_{22}(m_Z^2, m_s^2,m_s^2,m_b^2, m_Z^2,m_s^2)
+C_{23} (m_Z^2, m_s^2,m_s^2,m_b^2, m_Z^2,m_s^2)\biggr)\biggr ] \;.
\eea
Using the quark masses (in GeV) as $m_s=0.15$ and $m_b=4.4$,
$\sin^2 \theta_W=0.23$, $\alpha=1/127$, and the mass and width of $Z$
boson from
\cite{pdg}, we obtain the branching ratio including
one-loop corrections as
\be
{\cal B}(Z \to b \bar s)=4.08 \times 10^{-7}\;.
\ee
Since the expected sensitivity of giga-$Z$ collider is of the order
of $10^{-8}$ (which is at the level of SM expectation),
we emphasize that new physics effect could be detectable
in the rare decay $Z \to b \bar s$, if indeed  it affects
this mode.

Here, we have analyzed the rare decay modes $B_s \to l^+ l^-$ and
$B_s \to l^+ l^- \gamma$  which are mediated by
the $ b \to s $ FCNC transitions. We have considered the model
which induces tree level FCNC
coupling of $Z$ boson, due to the addition of
an extra vector like down quark to the matter sector. We found that
the branching ratios for the radiave leptonic decay modes $B_s \to
\mu^+ \mu^- \gamma, ~ (\tau^+ \tau^- \gamma)$ are
of the order of $10^{-8}~(10^{-7})$ which could be observable in the
Tevatron Run II. Furthermore, the branching ratio of
the pure leptonic mode  $B_s \to \tau^+ \tau^-$ found to be ${\cal O}(10
^{-6})$. This mode can be observed in the currently running $B$ factories
with improved $\tau$ tagging efficiency.

We have also analyzed the flavor changing decay of $Z$-boson to a
pair of down quarks $ Z \to b \bar s$. This $Z$ decay channel
may prove useful in searching for new flavor physics
beyond the SM at the TESLA or any other future
collider which may be designed to run at the $Z$-pole with high
luminosities, thus accumulating more than $10^9$ on-shell $Z$
bosons. With improved $b$-tagging efficiencies, the flavor
changing decay $Z \to b \bar s$ is the most likely
and the easiest one to detect
among the flavor changing hadronic $Z$ decays. It may be accessible
to the Giga-Z option even for branching ratio as small
as ${\cal B}(Z \to b \bar s) \sim 10^{-7}-10^{-6}$.

To conclude, the standard model results of the rare decays 
studied here which are
induced by FCNC transitions, are very small and cannot be
detected in the current
or near future experiments. These rare decays provide very sensitive
probe of new physics beyond the SM. Detection of these decays at
visible levels by any of the future colliders would be a clear
evidence of new physics.

{\bf{Acknowledgements}}
This work is
partly supported by the Department of Science and Technology,
Government of India, through Grant No. SR/FTP/PS-50/2001.

\appendix

\section{One-loop form factors }
The two-point and three-point one-loop form factors which are
defined as
\bea
C_0;&& C_\mu;\; C_{\mu \nu}(m_1^2, m_2^2, m_3^2, p_1^2, p_2^2,p_3^2)\nn\\
&&\equiv \int \frac{d^4 q}{i \pi^2}~ \frac{1;~ q_{\mu};~ q_\mu q_\nu}
{[q^2-m_1^2][(q+p_1)^2-m_2^2][(q-p_3)^2-m_3^2]}
\eea
\be
\tilde C_0; \tilde C_{\mu \nu} (m_1^2, m_2^2, m_3^2, p_1^2, p_2^2,p_3^2)
\equiv \int \frac{d^4 q}{i \pi^2} ~\frac{q^2;~q^2q_\mu q_\nu}
{[q^2-m_1^2][(q+p_1)^2-m_2^2][(q-p_3)^2-m_3^2]}\;,
\ee
where $\sum_ip_i=0 $ is to be understood above.

\be
B_0;~B_\mu(m_1^2,m_2^2,p^2) \equiv
\int \frac{d^4 q}{i \pi^2} \frac{1;~q_\mu}
{[q^2-m_1^2][(q+p)^2-m_2^2]}\;,
\ee

The coefficients $B_j$ with $j \in 0,1$, $C_j$ with
$j \in 0,11,12,21,22,23,24  $ are defined through the following
relation
\bea
&&B_\mu = p_\mu B_1\nn\\
&&C_\mu = p_{1 \mu}C_{11}+p_{2 \mu}+p_{2 \mu}C_{12}\nn\\
&&C_{\mu \nu}=p_{1 \mu}p_{1 \nu}C_{21}+p_{2 \mu}p_{2 \nu}C_{22}
+\{p_1~p_2\}_{\mu \nu}C_{23}+g_{\mu \nu}C_{24}\nn\\
&&\tilde C_{\mu \nu}=p_{1 \mu}p_{1 \nu}\tilde C_{21}+
p_{2 \mu}p_{2 \nu}\tilde C_{22}
+\{p_1~p_2\}_{\mu \nu}\tilde C_{23}+g_{\mu \nu}\tilde C_{24}
\eea
where $\{a~ b \}=a_\mu b_\nu+a_\nu b_\mu$.

%**************************************************************************

\end{document}